\documentclass[a4paper,aps,prd,12pt,preprintnumbers,showpacs,onecolumn,groupedaddress,superscriptaddress,nofootinbib,amsmath,amssymb]{revtex4-1}
\usepackage{graphicx}
\usepackage[utf8]{inputenc}
\usepackage[T1]{fontenc}
\usepackage{cmap}
\usepackage[normalem]{ulem}

\begin{document}
\linespread{1.667}
%\doublespacing

% Honorable mention of the Gravity Research Foundation 2023 Awards for Essays on Gravitation
\centerline{Honorable mention of the Gravity Research Foundation 2023 Awards for Essays on Gravitation}

\title{The sound of the event horizon}
\author{R. A. Konoplya}\email{roman.konoplya@gmail.com}
\affiliation{Research Centre for Theoretical Physics and Astrophysics, Institute of Physics, Silesian University in Opava, Bezručovo náměstí 13, CZ-74601 Opava, Czech Republic}
\date{March 29, 2023.}
%\pacs{03.65.Pm,04.30.-w,04.70.Bw}

\begin{abstract}
During the ringdown phase of a gravitational signal emitted by a black hole, the least damped quasinormal frequency dominates. If modifications to Einstein's theory induce noticeable deformations of the black-hole geometry only near the event horizon, the fundamental mode remains largely unaffected. However, even a small change near the event horizon can significantly impact the first few overtones, providing a means to probe the geometry of the event horizon. Overtones are stable against small deformations of spacetime at a distance from the black hole, allowing the event horizon to be distinguished from the surrounding environment. In contrast to echoes, overtones make a much larger energy contribution. These findings open up new avenues for future observations.
\end{abstract}

\maketitle

Quasinormal modes are a key feature of black holes, as they can be detected through gravitational waves \cite{LIGOScientific:2016aoc,LIGOScientific:2017vwq,LIGOScientific:2020zkf}. They depend only on the parameters of the black hole, making them an intrinsic characteristic of black holes. Quasinormal modes provide a window into the behavior of gravity in the strong-field regime, where the effects of general relativity are most pronounced \cite{Konoplya:2011qq,Berti:2009kk,Kokkotas:1999bd,Nollert:1999ji}.

However, it is widely believed that quasinormal modes alone cannot determine the geometry of the event horizon. Even a vastly different object, such as a wormhole, could mimic the ringdown of a Schwarzschild/Kerr black hole \cite{Damour:2007ap}. As a result, there would be no potentially observable imprint, unless one considers elusive echoes at asymptotically late times when the signal is strongly damped \cite{Cardoso:2016rao}. In this context, we will argue that, unlike the fundamental mode, the first few overtones may potentially serve as probes of the event horizon.

When perturbing Schwarzschild spacetime, the linearized Einstein equations can be simplified into the Schrödinger wavelike equation \cite{Regge:1957td,Zerilli:1970se}
\begin{equation}\label{Wave-like-equation}%
\left(\frac{d^2}{dr_*^2} + \omega^2 - V(r_*, \ell, M)\right)\Phi(r_*) = 0,
\end{equation}
with respect to the tortoise coordinate,
\begin{equation}\label{tortoise-coordinate}
dr_* \equiv \frac{dr}{f(r)}, \qquad f(r) = 1- \frac{2 M}{r}.
\end{equation}
Here $V(r_*, \ell, M)$ is the effective potential surrounding a black hole. The parameters $\ell$ and $M$ denote the multipole number and the mass of the black hole, respectively.
The proper oscillation frequencies of a perturbed black hole, known as \textit{quasinormal modes} \cite{Konoplya:2011qq,Berti:2009kk,Kokkotas:1999bd,Nollert:1999ji}, are eigenvalues of the above equation that satisfy special boundary conditions. Specifically, purely in-going waves at the event horizon and purely out-going waves at infinity. There have been many attempts to verify gravitational theory via compact objects, in both the weak (post-Newtonian) and strong gravity regimes, and thus far, no contradictions with Einstein's theory have been found \cite{LIGOScientific:2016aoc,LIGOScientific:2017vwq,LIGOScientific:2020zkf}. Therefore, it is natural to expect that deviations from Schwarzschild/Kerr geometry will only be observed in a small region near the event horizon.

\begin{figure}
\resizebox{\linewidth}{!}{\includegraphics{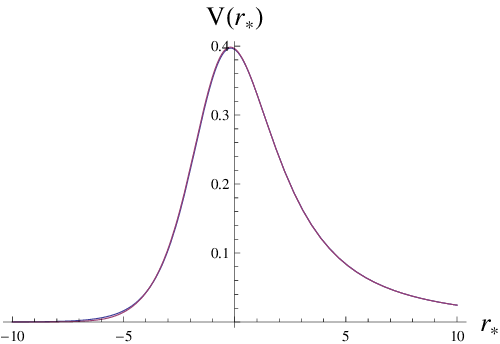}\includegraphics{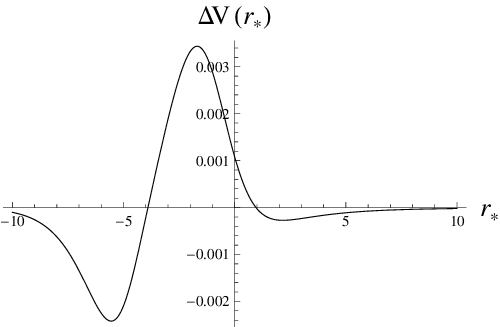}}
\caption{Left panel: Effective potentials of a scalar field ($\ell=1$, $r_0=2M=1$) for the  black hole deformed near the event horizon (blue) and Schwarzschild black hole (red). Right panel: difference between the potentials.}\label{fig:scalarpot}
\end{figure}

\begin{figure*}
\resizebox{\linewidth}{!}{\includegraphics{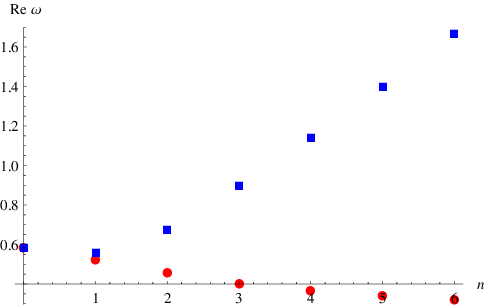}\includegraphics{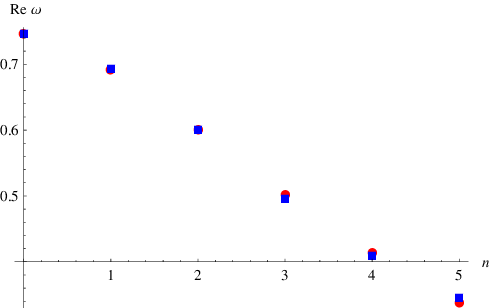}}
\caption{Left: The fundamental mode and first five overtones for the scalar field perturbations of the Schwarzschild black hole (red) and the black hole deformed in the near horizon zone (blue) as in fig. \ref{fig:scalarpot}. Right: The first five modes of the gravitational Regge-Wheeler potential (red) and the potential deformed at a distance from the black hole (blue): $\delta=0.006$ and $r_m=20 r_0$.}\label{fig:overtones}
\end{figure*}

As a result, the Schwarzschild metric function $f(r)$ and the effective potential undergo small augmentations
\begin{equation}\label{tortoise-coordinate}
f(r) \rightarrow f(r) + \Delta f(r),
\end{equation}
\begin{equation}\label{tortoise-coordinate}
V(r) \rightarrow V(r) + \Delta V(r).
\end{equation}
These augmentations could be due to extra-dimensional scenarios, quantum corrections, or alternative theories of gravity with small couplings, and the corresponding corrections $\Delta f(r)$ and $\Delta V(r)$ must decay quickly with distance. From a technical standpoint, the spectra of test fields, such as a scalar field, can be used for illustration purposes (as shown in Fig. \ref{fig:scalarpot}), without the need to specify the underlying gravitational theory. Alternatively, we can also consider deformations of the Regge-Wheeler effective potential directly (as depicted in Fig. \ref{fig:overtones}).

The ringdown phase of a gravitational wave signal emitted by a black hole is dominated by the longest-lived, fundamental mode. Typically, this mode is determined by the scattering near the peak of the effective potential $V(r)$ \cite{Mashhoon:1982im,Schutz:1985km}, which is located at some distance from the black hole (at $r = 3 M$ for the Schwarzschild case). Deformations $\Delta f(r)$ and $\Delta V(r)$ are expected to be negligibly small at such a separation from the horizon (Fig. \ref{fig:scalarpot}), so the fundamental mode would remain practically unaltered.

However, we will show that the first few overtones are very sensitive to even the slightest near-horizon changes, thereby providing information about the event horizon. The idea that overtones may be related to near-horizon behavior is not entirely unexpected, as in the limit of an infinitely high overtone number $n$, it can be analytically proven that the event horizon determines the asymptotic behavior of highly damped quasinormal modes \cite{Nollert:1993zz,Motl:2002hd,Motl:2003cd}. The first work to show that modes might be highly sensitive to small deformations of the effective potential everywhere outside the black hole was done by H.-P. Nollert, who turned the Regge-Wheeler potential into a series of step potentials. Instead of reproducing the Schwarzschild spectrum, he observed the aforementioned sensitivity of modes to even small changes in the potential \cite{Nollert:1996rf}. The same was done for a smooth sinusoidal deformation of the potential \cite{Jaramillo:2021tmt} in the whole space. It was observed that the fundamental mode practically does not change once the amplitude of the deformation is small enough, but the overtones explode.

Here we demonstrate the importance of the {\it location} of small deformations in the geometry on the sensitivity of overtones. Specifically, if a deformation is made only in a small region near the event horizon, causing the deformed potential to coincide with the Schwarzschild potential near the peak, then an outburst of overtones occurs, while the fundamental modes remain unaffected.

Figure \ref{fig:scalarpot} illustrates an example of such a small deformation of the effective potential. The Schwarzschild and deformed potentials are nearly identical and cannot be visually distinguished. The amplitude of the near horizon deformation, as shown in the right plot of Fig. 1, is a few orders smaller than the height of the main peak. Calculation of several lowest quasinormal frequencies using two independent, accurate, and convergent methods (Leaver continued fraction method and Bernstein polynomial approach) reveals that while the fundamental mode changes by only $0.4 \%$, the first overtone already changes by about $6 \%$, and the fifth overtone may deviate from its Schwarzschild limit by hundreds of percent \cite{Konoplya:2022pbc}. The difference is mainly in the real oscillation frequency given by $Re(\omega)$, as shown in Fig. \ref{fig:overtones}.

Meanwhile, overtones are stable against small deformations of the geometry at a distance from the black hole. Thus, the contribution due to the event horizon can always be distinguished from that of the environment, opening a new aspect for future observations. To prove this, we consider a simple augmentation of the Regge-Wheeler potential at a distance from the black hole, implying possible effect of the environment. The augmentation  approaches zero at the horizon as $\Delta V\propto(r-r_0)^h$ and at infinity as $\Delta V\propto r^{-a}$, having the maximum value $\Delta V_{max}=\delta/r_0^2$ at $r=r_m>r_0$, where $r_{0}$ is the radius of the event horizon.

To illustrate this concept, we consider $\delta=0.006$ and $r_m=20 r_0$, which is two orders smaller than the height of the main Schwarzschild peak for $\ell=2$ and much larger than a realistic astrophysical environment, such as an accretion disk, could be. We resort to such astrophysically large deformations to be able to plot at least small deviations of the overtones. Such an astrophysically big deformation of the potential leads to very small corrections of the first several overtones, as shown on the right plot of Fig. \ref{fig:overtones}. While the first few modes change by a small fraction of one percent, the fourth overtone deviates from its Schwarzschild value by only $2 \%$. For astrophysically relevant values of $\delta$ and $r_m$, the latter, normally, must be larger than the innermost stable circular orbit, the first several overtones coincide with their Schwarzschild values with high accuracy. The corrections become even smaller if we shift the deformation farther from the black hole. The sensitivity of overtones is not significantly affected by the specific shape of the deformation function at a distance.  Thus, we conclude that the phenomenon of the strong change of overtones  indeed happens due to deformations near the horizon, and no realistic astrophysical environment could produce such a noticeable outburst.

The question remains: can the overtones that probe the near-horizon geometry be observed, at least in principle? According to the analysis in \cite{Cotesta:2022pci}, no overtones could be distinguished in the signal of $GW150914$ post-merger data. However, two aspects suggest a positive answer. Firstly, the deformations localized solely near the event horizon need not be small, as it happens in some alternative theories of gravity. In these cases, even the first overtone may change so strongly that it becomes concurrent with the fundamental mode \cite{Konoplya:2022hll}. Secondly, reproducing the earlier stage of the ringdown phase requires not only the fundamental mode, but also quite a few overtones \cite{Isi:2019aib}. This means that the contribution of overtones could be extracted from the time domain profile \cite{Gundlach:1993tp} at the beginning of the ringdown. Moreover, as shown in Fig. 3 of \cite{Konoplya:2022pbc}, the energy of excitation of overtones is definitely many orders larger than that of the asymptotically late-time modification of the signal - echoes \cite{Cardoso:2016rao}. After all, there are indications that future experiments with the Laser Interferometer Space Antenna (LISA) \cite{Oshita:2022yry} may provide an opportunity to observe overtones. Thus, while current observations cannot reveal the near-horizon geometry of black holes, there is hope that it may be possible in the future via analysis of the few lowest overtones.

\begin{acknowledgments}
The author acknowledges useful discussions with Alexander Zhidenko.
\end{acknowledgments}


\begin{thebibliography}{80}

%\cite{LIGOScientific:2016aoc}
\bibitem{LIGOScientific:2016aoc}
B.~P.~Abbott \textit{et al.} [LIGO Scientific and Virgo],
%``Observation of Gravitational Waves from a Binary Black Hole Merger,''
Phys. Rev. Lett. \textbf{116}, no.6, 061102 (2016)
%doi:10.1103/PhysRevLett.116.061102
%[arXiv:1602.03837 [gr-qc]].

%\cite{LIGOScientific:2017vwq}
\bibitem{LIGOScientific:2017vwq}
B.~P.~Abbott \textit{et al.} [LIGO Scientific and Virgo],
%``GW170817: Observation of Gravitational Waves from a Binary Neutron Star Inspiral,''
Phys. Rev. Lett. \textbf{119}, no.16, 161101 (2017)
%doi:10.1103/PhysRevLett.119.161101
%[arXiv:1710.05832 [gr-qc]].

%\cite{LIGOScientific:2020zkf}
\bibitem{LIGOScientific:2020zkf}
R.~Abbott \textit{et al.} [LIGO Scientific and Virgo],
%``GW190814: Gravitational Waves from the Coalescence of a 23 Solar Mass Black Hole with a 2.6 Solar Mass Compact Object,''
Astrophys. J. Lett. \textbf{896}, no.2, L44 (2020)
%doi:10.3847/2041-8213/ab960f
%[arXiv:2006.12611 [astro-ph.HE]].


%\bibitem{reviews}
%\cite{Konoplya:2011qq}
\bibitem{Konoplya:2011qq}
R.~A.~Konoplya and A.~Zhidenko,
%``Quasinormal modes of black holes: From astrophysics to string theory,''
Rev.\ Mod.\ Phys.\  {\bf 83}, 793 (2011);
doi:10.1103/RevModPhys.83.793
[arXiv:1102.4014 [gr-qc]];
%\cite{Berti:2009kk}
\bibitem{Berti:2009kk}
E.~Berti, V.~Cardoso and A.~O.~Starinets,
%``Quasinormal modes of black holes and black branes,''
Class.\ Quant.\ Grav.\  {\bf 26}, 163001 (2009);
%doi:10.1088/0264-9381/26/16/163001
%[arXiv:0905.2975 [gr-qc]];\\
%\cite{Kokkotas:1999bd}
\bibitem{Kokkotas:1999bd}
K.~D.~Kokkotas and B.~G.~Schmidt,
%``Quasinormal modes of stars and black holes,''
Living Rev.\ Rel.\  {\bf 2}, 2 (1999);
%doi:10.12942/lrr-1999-2
%[gr-qc/9909058];\\
%\cite{Nollert:1999ji}
\bibitem{Nollert:1999ji}
H.~P.~Nollert,
%``TOPICAL REVIEW: Quasinormal modes: the characteristic `sound' of black holes and neutron stars,''
Class.\ Quant.\ Grav.\  {\bf 16}, R159 (1999).
%doi:10.1088/0264-9381/16/12/201

%\cite{Damour:2007ap}
\bibitem{Damour:2007ap}
T.~Damour and S.~N.~Solodukhin,
%``Wormholes as black hole foils,''
Phys. Rev. D \textbf{76}, 024016 (2007)
%doi:10.1103/PhysRevD.76.024016
%[arXiv:0704.2667 [gr-qc]].

%\cite{Cardoso:2016rao}
\bibitem{Cardoso:2016rao}
V.~Cardoso, E.~Franzin and P.~Pani,
%``Is the gravitational-wave ringdown a probe of the event horizon?,''
Phys. Rev. Lett. \textbf{116}, no.17, 171101 (2016)
[erratum: Phys. Rev. Lett. \textbf{117}, no.8, 089902 (2016)]
%doi:10.1103/PhysRevLett.116.171101
%[arXiv:1602.07309 [gr-qc]].


%\cite{Regge:1957td}
\bibitem{Regge:1957td}
T.~Regge and J.~A.~Wheeler,
%``Stability of a Schwarzschild singularity,''
Phys. Rev. \textbf{108}, 1063-1069 (1957)

%\cite{Zerilli:1970se}
\bibitem{Zerilli:1970se}
F.~J.~Zerilli,
%``Effective potential for even parity Regge-Wheeler gravitational perturbation equations,''
Phys. Rev. Lett. \textbf{24}, 737-738 (1970)



%\cite{Mashhoon:1982im}
\bibitem{Mashhoon:1982im}
B.~Mashhoon, ``Quasinormal modes of a black hole,'' Contribution to: 3rd Marcel Grossmann Meeting on the Recent Developments of General Relativity, 599-608 (1982)

%\cite{Schutz:1985km}
\bibitem{Schutz:1985km}
B.~F.~Schutz and C.~M.~Will,
%``BLACK HOLE NORMAL MODES: A SEMIANALYTIC APPROACH,''
Astrophys. J. Lett. \textbf{291}, L33-L36 (1985)


%\cite{Nollert:1993zz}
\bibitem{Nollert:1993zz}
H.~P.~Nollert,
%``Quasinormal modes of Schwarzschild black holes: The determination of quasinormal frequencies with very large imaginary parts,''
Phys. Rev. D \textbf{47}, 5253-5258 (1993)


%\cite{Motl:2002hd}
\bibitem{Motl:2002hd}
L.~Motl,
%``An Analytical computation of asymptotic Schwarzschild quasinormal frequencies,''
Adv. Theor. Math. Phys. \textbf{6}, 1135-1162 (2003)
%doi:10.4310/ATMP.2002.v6.n6.a3
%[arXiv:gr-qc/0212096 [gr-qc]].

%\cite{Motl:2003cd}
\bibitem{Motl:2003cd}
L.~Motl and A.~Neitzke,
%``Asymptotic black hole quasinormal frequencies,''
Adv. Theor. Math. Phys. \textbf{7}, no.2, 307-330 (2003)
%doi:10.4310/ATMP.2003.v7.n2.a4
%[arXiv:hep-th/0301173 [hep-th]].




%\cite{Nollert:1996rf}
\bibitem{Nollert:1996rf}
H.~P.~Nollert,
%``About the significance of quasinormal modes of black holes,''
Phys. Rev. D \textbf{53}, 4397-4402 (1996)
%doi:10.1103/PhysRevD.53.4397
%[arXiv:gr-qc/9602032 [gr-qc]].


%\cite{Jaramillo:2021tmt}
\bibitem{Jaramillo:2021tmt}
J.~L.~Jaramillo, R.~Panosso Macedo and L.~A.~Sheikh,
%``Gravitational Wave Signatures of Black Hole Quasinormal Mode Instability,''
Phys. Rev. Lett. \textbf{128}, no.21, 211102 (2022)
%doi:10.1103/PhysRevLett.128.211102
%[arXiv:2105.03451 [gr-qc]].

%\cite{Konoplya:2022pbc}
\bibitem{Konoplya:2022pbc}
R.~A.~Konoplya and A.~Zhidenko,
%``First few overtones probe the event horizon geometry,''
[arXiv:2209.00679 [gr-qc]].

%\cite{Cotesta:2022pci}
\bibitem{Cotesta:2022pci}
R.~Cotesta, G.~Carullo, E.~Berti and V.~Cardoso,
%``Analysis of Ringdown Overtones in GW150914,''
Phys. Rev. Lett. \textbf{129}, no.11, 111102 (2022)
%doi:10.1103/PhysRevLett.129.111102
%[arXiv:2201.00822 [gr-qc]].





%\cite{Konoplya:2022hll}
\bibitem{Konoplya:2022hll}
R.~A.~Konoplya, A.~F.~Zinhailo, J.~Kunz, Z.~Stuchlik and A.~Zhidenko,
%``Quasinormal ringing of regular black holes in asymptotically safe gravity: the importance of overtones,''
JCAP \textbf{10}, 091 (2022)
%doi:10.1088/1475-7516/2022/10/091
[arXiv:2206.14714 [gr-qc]].






%\cite{Isi:2019aib}
\bibitem{Isi:2019aib}
M.~Isi, M.~Giesler, W.~M.~Farr, M.~A.~Scheel and S.~A.~Teukolsky,
%``Testing the no-hair theorem with GW150914,''
Phys. Rev. Lett. \textbf{123}, no.11, 111102 (2019)
%
%doi:10.1103/PhysRevLett.123.111102
%[arXiv:1905.00869 [gr-qc]].

%\cite{Gundlach:1993tp}
\bibitem{Gundlach:1993tp}
C.~Gundlach, R.~H.~Price and J.~Pullin,
%``Late time behavior of stellar collapse and explosions: 1. Linearized perturbations,''
Phys. Rev. D \textbf{49}, 883-889 (1994)
%doi:10.1103/PhysRevD.49.883
%[arXiv:gr-qc/9307009 [gr-qc]].





%\cite{Oshita:2022yry}
\bibitem{Oshita:2022yry}
N.~Oshita and D.~Tsuna,
%``Slowly Decaying Ringdown of a Rapidly Spinning Black Hole: Probing the No-Hair Theorem by Small Mass-Ratio Mergers with LISA,''
[arXiv:2210.14049 [gr-qc]].

\end{thebibliography}
\end{document}